# AI in Education: Rationale, Principles, and Instructional Implications

Eyvind Elstad[1]

University of Oslo

December 2nd 2024

**Abstract**

This study explores the integration of generative artificial intelligence (AI) in school settings, evaluating both its potential benefits and the risks it poses to education. The expanding use of AI by students necessitates a critical examination of its impact on learning processes and pedagogical practices. Generative AI, especially large language models (LLMs) like ChatGPT, have the capacity to simulate human-like text, music, movies et cetera raising important questions about their role in education. This article reviews the theoretical foundations of intelligence, creativity, and AI, differentiating LLMs from traditional search engines and emphasizing the need for critical source competence among students. Empirical evidence on the effects of AI in classrooms is currently limited. However, AI's ability to provide personalized learning support, facilitate complex problem-solving, and offer educational platforms presents both opportunities and challenges. While AI can assist in producing text and enhancing productivity, concerns arise over its potential to undermine deep learning and critical thinking when students use it as a shortcut. The study highlights the importance of deliberate educational strategies to ensure that AI supplements rather than replaces genuine cognitive effort. By examining cognitive theories and the ecological framework of classrooms, the study underscores the importance of integrating AI in a way that fosters active and meaningful learning. AI's role should be context-dependent, guided by the pedagogical rationale specific to different educational stages and subjects. The study concludes with practical inferences for teachers on utilizing AI effectively, emphasizing promoting learning with understanding, and fostering students' critical engagement with AI-generated content. We need a balanced and informed approach to leveraging AI in education, prioritizing the long-term development of students' knowledge, skills, and critical competencies.

## Introduction

The use of artificial intelligence (AI) in schools is a topic that is increasingly engaging many stakeholders. This study provides the reader with a foundation to consider how generative AI can be applied in teaching various school subjects and to identify classroom settings in which the technology should not be used. There is a growing tendency for students to use AI in their schoolwork. Students chart their own course, and they do so regardless of what a given teacher thinks about how students should use technology. Young people are both the most frequent users of AI and those who believe to the greatest extent that the information they obtain from AI services is correct and credible. This latter point highlights the school's duty to facilitate students' development of critical source competence. As teachers, school leaders, and educational authorities, we must consider not only how schools should manage students' independent use of AI but also when and how AI could be included in teaching.

---

[1] Email: eyvind.elstad@ils.uio.no

Finding solutions to recognized problems is often connected with human creativity, which is based partially or even entirely on intelligence (Sternberg, 1999). The term 'intelligence' refers to the human ability to find solutions to new tasks using reasoning (Cattell, 1963), but technology can interact with human intelligence and creativity when problems need to be solved (Jia et al., 2024). AI refers to how computer systems can perform tasks that normally require – or at least once required – human intelligence. For example, AI can be useful for creative processes (Doshi & Hauser, 2024) and for increased productivity, as when it helps workers in text-intensive occupations compose their materials (Noy & Zhang, 2023). Traditionally, humans have been the central actors in education. But through the emergence of generative AI based on large language models (LLMs), this reality is being challenged (Staneva & Elliott, 2023): for the first time in history, a non-human actor exists that can use human language in a way that resembles human communication so much that it can be hard to distinguish between a person and a machine. This raises many challenges and risks for the school and society in general, but it also offers new opportunities. Today's teachers find themselves amid these challenges.

Learning is a long-term activity. The education in secondary school is built on the fundamental knowledge and skills acquired in primary school and aims to further develop students' knowledge, critical thinking, and specialized skills as – for some learners – a preparation for higher education. In principle, the use of AI in preparatory programs should above all have a pedagogical rationale. Vocational training, on the other hand, aims to provide the learning individual with a profession-specific competence that provides a basis for becoming a professional practitioner. Vocational educations (professional educations, specialized college educations, and vocational education programs) should embrace the opportunities that AI provides in the professional operations to which these educational paths lead. The reason is that technological development makes those occupational tasks easier, more efficient, and effective, and AI has the potential to promote workplace productivity in many areas.

AI is the simulation of intelligence in machines (Cukurova, 2024); it receives impressions and impulses from its environment and produces an intelligent way of being (Russel & Norvig, 2020). Its manifestations can include translation from one language to another, creating images and videos, recognizing speech, aiding in decision-making and much more (Epstein et al., 2023). When data systems are capable of learning from their own experiences and solving complex problems, we speak of AI. In this way, the technology may appear to be intelligent, but it is still artificial.

An LLM is a specialized type of AI that has been trained on large amounts of text to 'understand' existing content and generate new content. It is important to recognize that LLMs are neither search engines that seek facts nor databases; rather, they are pattern-seeking engines that predict the next word in a sequence of words. LLMs do not check facts, and they do not find facts. The foundation of a LLM lies in a variant of a computer program that has the capacity to produce text. These models – referred to as generative AI - have learned by training on large amounts of textual data sources. The most widely known one is ChatGPT, which launched a free version on November 30, 2022. Since its launch, ChatGPT and its performance have received substantial attention.

GPT is an acronym for the generative, pre-trained and transformer, each of which is explained below (Wolfram, 2023). Generative refers to the tool's ability to produce new text based on text that has been fed into it. Students should be critical and able to evaluate information using their own knowledge. ChatGPT bases its responses on large amounts of data that may contain errors or even fabrications. It simulates human answers using probability and can therefore produce inaccurate or fanciful; when this happens, the technology is said to 'hallucinate'.

Being pre-trained means that ChatGPT has been prepared on a large volume of actual text and enables it to generate coherent and contextually relevant text based on the inputs the user supplies. When ChatGPT was launched, users quickly became aware that it had been pre-trained on limited material. But this has changed over time, and it functions much better today (Yuan et al., 2024).

The transformer aspect refers to a network architecture that can handle sequences of data such as words. It was a professional community at Google that developed a transformer (Vaswani et al., 2017), which was first tested out in a translation program. It appears to 'understand' the context around words and sentences better than its predecessors in this network architecture. Imagine asking ChatGPT to translate this sentence into German: 'I arrived at the bank after crossing the road' (Uszkoreit, 2017). The correct translation is: ' Ich kam zur Bank, nachdem ich die Straße überquert hatte'. But ChatGPT gets confused when you ask it to translate 'I arrived at the bank after crossing the river' because the word 'bank' in English can mean a financial institution or the land next to a river, depending on context. In the sentence 'I arrived at the bank after crossing the river', if 'bank' is meant to be 'riverbank', the correct translation to German would be: 'Ich kam ans Flussufer, nachdem ich den Fluss überquert hatte'.

This means that ChatGPT's attention mechanism enables it to figure out that the word 'bank' has different meanings in the two sentences. The transformer property consists of an attention mechanism that allows the LLM to weigh the significance of different parts of the sentence that was fed in and to process all parts of the sentence simultaneously: when 'the river' occurs at the end of the sentence, the meaning content in 'bank' (which appears before 'river') shifts: it has attention.

Although pre-training gives the language model a general understanding of language, ChatGPT has also been supplemented with fine-tuning that often creates positive framings for the answers it supplies. This is shown by asking ChatGPT the following three questions: 'Can I be proud of being' (1) queer, (2) a woman and (3) a man? It answers the first two as follows: 'Absolutely! Feeling pride in being [queer/a woman] is a valid and empowering experience'. To the question 'Can I be proud of being a man?,' it answers at length:

> It's important to consider the societal context in which such pride is expressed. Historically, expressions of male pride have been tied to patriarchal structures and the marginalization of women and persons who do not identify with a binary gender. As a result, discussions about male pride can be sensitive and complex.

What does this mean? ChatGPT and other LLMs do not provide neutral answers. Language models like ChatGPT, Gemini, Llama, and others simply have their own political slant (Feng et al., 2023). This implies that those using these LLMs must have the critical judgment needed to evaluate the answers they provide. However, a chatbot can also curb erroneous conclusions (Costello et al., 2014). These conflicting consequences have implications for teachers' work.

The type of AI based on LLMs has been described as 'a stochastic parrot' (Cukurova, 2024). The word 'stochastic' indicates that the process leading to a result has a random component. For example, when the AI model generates a text, the model calculates probabilities for the word or expression that should come next and often tests these probabilities, leading to variations in outputs even when the same prompt is used multiple times. The parrot element refers to the model repeating words or sentences it has heard, which happens without the model 'understanding' these words or sentences. It should be noted that generative AI cannot, so far, truly understand the meaning behind the words that are fed into it. This means that the answers it generates can often be superficial and lack depth and insight (Bogust, 2022). Thus, the language model does not create original thoughts but so far reproduces human-like text based on the patterns it learned during training on a large dataset of text. In other words, the language model does not understand text in the way humans do but can mimic coherent and contextually appropriate linguistic responses. Therefore, it is among the class of language models that can be characterized as stochastic parrots. This points out one of the limitations of the current language models.

Whether we are facing a new turn in the development of AI – an artificial general intelligence that resembles human intelligence (Mitchell, 2024) – remains an open question. Such an intelligence is expected to have the ability (rather like the human form) to use its 'intellect' to solve a wide range of problems by transferring knowledge from one situation to another (Sonko et al., 2024). This form of intelligence is also often referred to as strong AI, and some believe it is possible to develop such artificial general intelligence. If this happens, society may change considerably (Bengio et al., 2024), but it is not clear whether it is possible to develop such an intelligence and, if so, when that technology will be available (Mitchell, 2024). At the time of writing (November 2024), there is no consensus that general AI exists, but some see OpenAI's language models as a step towards an artificial general intelligence (Robison, 2024).

AI is not a new phenomenon; it has been a part of the development of technology for several decades. We are familiar with AI using various types of technological solutions: the self-propelled and pre-programmed robotic vacuum cleaner uses AI, translation programs such as Google Translate use AI, Microsoft Word uses AI to suggest linguistic improvements, Gmail uses AI to filter spam, and YouTube and Netflix use AI to suggest playlists. The list of AI applications is long. In addition, AI is expected to have impacts on several workplaces (including teaching), but as things now stand, it is difficult to have a clear opinion on how those impacts will be manifest in tomorrow's labour market (Staneva & Elliot, 2023).

Many are familiar with rule-driven AI using chatbots such as those that public agencies, banks and other private businesses have been using for several years. This rule-based AI only does what the machine is prepared to do. Things are different with generative AI based on LLMs. This article is about the use of generative AI in schools.

Interest in AI over time can show fluctuations between flourishing and declining interest (Toosi et al., 2021). In recent years, growing interest has been typical. The emergence of ChatGPT has received widespread attention, but there are several other chatbots in addition to ChatGPT: Gemini, Copilot, Claude, Perplexity, Jasper, Chatsonic, Socratic and so on.

## AI in Schools

Generative AI has been integrated into what are called educational platforms, with SchoolAI, Canvas, Mizou, Curious and Learnlab the best known. With these platforms, the teachers have their own dashboards, which are digital modules that enable teachers to gain an overview of how the students are progressing on the tasks they have been assigned. There are several support functions that can help students in certain phases of writing work (choosing a topic, finding material, etc.). As of this writing, we do not know what impact this might have on students' learning progress, but anecdotal reports indicate that teachers find such educational platforms meaningful for their work.

AI is in the process of creating a significant societal upheaval in many areas. In schools, discussions about generative AI have revolved around how using that technology can be both a resource and a problem for teaching and learning. AI has already affected students' learning processes and teachers' work in the relatively short period that ChatGPT have existed. The message is that the conditions for teachers' work have changed: they should use the possibly beneficial approaches that AI offers and avoid the disadvantageous ones.

So far, there is little research that uses empirical evidence to report how teachers can and should benefit from using AI, how and whether AI can be beneficial for the learning process, the academic topics for which students' use of chatbots is suitable, or for which grades AI is best. All teachers must rely on their professional judgment. As noted above, it is the rationale of education and the justification of school subjects that should be emphasized when the question of using AI arises.

What matters when choosing technology? When, how and why AI and other technological opportunities are relevant to the teacher's work is connected to the justification of education (Salomon, 2016). Generative AI is only the beginning of a development whose outcomes do not know. Children starting school in autumn 2025 will complete their basic education in the 2030s. Most students then take further education, and many continue into higher education. The first grader in 2025 might be imagined completing their master education in 2043 or later. It is difficult to imagine how society will be structured then.

The purpose of primary education revolves around the understanding of and reasoning for education at the primary level. It is not the use of technology itself that is important, but the pedagogical utility of that use. In today's schools, digital skills are often defined as basic skills, and that is not sufficient to meet future realities. Students must also have knowledge of technology. Skills presuppose broad knowledge that is acquired in subject teaching. In today's schools, computer science is not a separate subject. In the Danish public school system, a separate subject called 'Technology Comprehension' has been introduced. This is an example of a forward-looking renewal of school content. Something similar is also needed in other countries' education.

The purpose of generative AI in primary education and in preparatory programs for higher education should be to help schools cultivate long-term learning rather than simply having students use more technology in school. Age and maturity should matter. The big tech giants are developing chatbots targeting increasingly younger age groups. For example, Google has launched chatbots for students aged 13 and up. This means that the question of how schools should relate to AI will soon concern middle schools. Some have even argued that AI is also relevant in grades 1–7 (Gates, 2024), although that view is highly controversial.

Through higher education, the learning individual is supposed to develop advanced knowledge and skills. This can be part of general or professional education. Higher education is the third level in the education system; its purpose is to offer opportunities for advanced studies and professional development that serve both individuals and the broader society. In many professions, technological advancements have eased traditional workloads, and that could also become the case for teachers. However, it cannot be blithely assumed to apply to the education sector. There is a fundamental difference between technology that streamlines production (i.e., that creates productivity gains) and technology that is to support students' learning processes. In school, it is precisely learning that is the entire purpose of the effort, and it is not always the case that productivity gains stemming from technology use serve that goal. Schooling and further education aim to enrich students with deep-seated knowledge and skills that will last in the long run.

Teachers must use professional judgment in deciding of when generative AI should be used in the pedagogical effort. The central element in teaching situations lies in the teacher's decisions. Various factors influence these choices when it comes to selecting teaching methods, including context, the nature of the learning material and students' existing knowledge levels. This implies that it is impractical to propose all-encompassing guidelines for the use of AI in the classroom across all subjects and age groups. AI is also merely one of many technological alternatives that can be used in an educational context. Teachers inevitably make decisions under uncertain conditions, which requires the use of sound pedagogical judgment.

Teaching has been compared to an art form (James, 1899/1983). At its best, research helps us abstain from counterproductive practices, but becoming a skilled teacher requires intuition and a keen understanding. This ability to explain the learning material in an educationally beneficial way and to respond well to students' questions is at the core of the art of teaching.

Every teacher should strive to develop a distinctive set of effective and efficient teaching methods and a repertoire of content representations. This development comes through repeated teaching experiences using the available technological tools that teachers have at their disposal. AI is just one of these tools. A teaching repertoire refers to the range of methods, strategies, techniques and resources a teacher uses to instruct students or create structures for their engagement and school activities. A functioning teaching repertoire should include a variety of approaches to meet students' different learning needs. By developing a flexible teaching repertoire, teachers can influence not only their students' learning process but also establish a learning environment that gives students rich opportunities to take active control of their own learning efforts.

The very act of learning and the way learning is viewed can change because of changes to the learning environments brought about by AI. One change may occur through personalized learning experiences. All students can in principle have access to virtual teaching tailored to their individual needs, interests and learning pace. With AI technology available over the internet, information becomes more accessible to more people, regardless of geographic, economic or social background. This can help level the playing field in education. On the other hand, a school that emphasizes students' ability to manage their own learning processes can also reinforce differences in students' opportunities to succeed in their education.

The simplest way to understand the concept of learning is that it is about the process of acquiring new knowledge and a better understanding (higher order knowledge). But the term 'to acquire' can be misleading since new knowledge sprouts from old knowledge: indeed, we can say that the learning individual acquires knowledge while meaning is constructed in a personal way and based on what that individual already knows. The term meaning construction implies that learning should be an active process: the change that occurs is achieved through what the student does mentally and/or physically. This understanding of learning is based on a cognitive theoretical outlook (Anderson, 2020), but it is limited to cognitive processes in individuals. What is needed in this paper is a theory about students' interactions with technology. It is not controversial to note that technology contributes to learning by providing academic support. However, the internal processing and storage of knowledge in the brain is the core of the learning process, not the information stored on digital devices. The goal of this paper is to present a useful theoretical framework for understanding how people learn, both with and without the help of AI. To achieve such a holistic understanding requires expanding the area of focus beyond the individual and incorporating theories that recognize generative AI as a possible external resource for the learning process. This entails acknowledging that thinking can be distributed: that is, students' thought processes not only take place inside but also in interaction with digital devices (and of course other students and teachers). I explore cognitive theory, ecological theory and the theory of distributed cognition, on which I elaborate below. This theoretical framework is therefore not limited to the mental system inside an individual's brain but recognizes that thinking is an extended phenomenon that also includes tools such as digital devices, other people, and the very environment with which the individual interacts (Salomon, 1997). This perspective gives us a more holistic understanding of the learning process in a world where technology plays an increasingly major role.

Information from a Google search is different from what is generated by an AI tool like ChatGPT. Based on its responses from large datasets full of potential errors, ChatGPT simulates human responses using probability and can therefore produce false or fictional results; as noted above, they can hallucinate. Unlike real understanding, AI-driven responses often lack depth and insight. Because of this, students should be critical and able to evaluate information using their own knowledge, but to be critical, students must have such knowledge in the first place. Teachers must therefore develop strategies and assessments to ensure that students develop the necessary knowledge and understanding, regardless of access to AI tools. This requires educational planning to assist students in the learning process.

AI can affect the learning process in several ways. The concept of student + technology addresses the idea that technology can extend and enhance an individual's thinking ability when the two function together as a unit. On the positive side, technology can act as a cognitive support and enhance the learning process. It can help a person perform complex tasks or understand complex ideas that might not have otherwise been mastered. In such cases, humans and technology combine their capacities to form a more effective thinking unit. On the downside, the use of technology can weaken the learning process (Abbas et al., 2024; Bastani et al., 2024; Shine, 2024; Zhang & Noyes, 2023). If a student uses technology to copy answers directly without engaging cognitively or striving to understand the material, technology becomes a short cut that derails the learning process. This can result in an adequate external outcome, such as a well-written essay, that lacks any underlying learning or the acquisition of new knowledge. This is the greatest danger of AI in schools.

The adage 'no pain, no gain' helps emphasize the importance of effort in the learning process (Bjork & Bjork, 2020; Kirschner et al., 2022). Education must facilitate that effort through desirable difficulties. Although technology can increase productivity and improve performance, it must be emphasized that the student still needs to engage actively and cognitively to achieve learning with genuine understanding.

The mechanisms of student motivation are complex and must be tailored according to age level and maturity. While motivation is often considered crucial for getting students to learn, there is no direct causal relationship by which motivation alone leads to learning (Kirschner et al., 2022). Instead, the satisfaction of succeeding at learning tasks can stimulate further motivation. This is due to the physiological effects of a chemical released in the brain. The desire to learn does not necessarily come before the effort.

The content of school subjects is hierarchically organized, but students read a text line by line or listen to teachers and other students speak; it is a linear process. The challenge is to extract meaning from the content that exists hierarchically in messages that occur linearly (Willingham, 2023). This mental reconstruction requires concentration, effort, dedication and sometimes even mental pain: fatigue from thinking, frustration at not understanding something immediately or the challenge of overcoming sometimes daunting obstacles in the learning process (David et al., 2024). The gain is the improved understanding that comes from sustained mental effort. To endure the mental pain, motivation, resilience and perseverance are necessary. If a student cannot exercise self-discipline, tutors who provide a framework for the learning work that stimulates self-discipline are necessary. This can be called learning management.

In practice, learning management can involve designing a decision architecture based on the principle of libertarian paternalism (Elstad, 2008): learning environments must be designed so that students have incentives (cohesion and structure) to overcome the mental pain that effort gives. Paternalism comes into play with the teacher setting the framework for teaching: instructions, expectations and control. The libertarian part comes into play because students will have some autonomy and co-determination. The goal of libertarian paternalism is to influence each student's decisions in ways that will make them better off through perseverance while preserving choice. Perseverance must be learned; it is not innate.

The driving force in perseverance has an internal and an external component, and it is the interplay between them that matters. Incentives can be built into the plan for learning work, support functions that may be needed and a mutual understanding of what the product of the effort involves (e.g., assessment criteria). Learning management can thus influence student decisions; the learning environment should drive students to do more than what is in their comfort zone. If students need peace and order to concentrate on the learning work, this is referred to as classroom management. Later in this paper, ecological theory is discussed, as it is precisely a theory about the interactions between teacher, students and technology in line with these premises.

Will the introduction of generative AI into society signal a more radical change in the theoretical framework of educational research that starts with the human being as the fundamental premise for learning in schools? We do not know. Do we need a more comprehensive theoretical shift in education research that reflects our current digital reality and the influence of AI? We do not know. For the first time in history, a non-human actor capable of using human language in a way so like human usage that it challenges the boundaries of what human communication and what is machine communication is emerging (Krumsvik, 2023). In examining this question, a sharp distinction must be made between vocational education that qualifies people for specific professional tasks on the one hand and education that lays the competence foundation for further academic education on the other.

For vocational trainings where the tasks are significantly changed due to AI, we need to establish a theoretical perspective that focuses on the performance that machine and human combine to provide. In some areas, operations can be fully or partially automated while the worker's role becomes to, for example, initiate machine performance, handle extraordinary situations and oversee everything. In other areas, cognition can be distributed across individuals and systems of artefacts (e.g., navigation maps, computation-based devices and communication systems; see Hutchins, 2020). Our theoretical frameworks and the interactions between humans and machines (including AI) must reflect this complexity.

An example of such a framework is the theory of computer-supported cooperative work (Chen et al., 2018; Janssen & Kirschner, 2020). What underlies this term is a group of theories intended to capture how humans collaborate using computer systems such as AI in autonomous vehicles, at a nuclear power plant or in the electronic systems of a jet aircraft and how these systems can be designed to support collaboration effectively (Cope et al., 2021; Howard, 2019).

In educational contexts that lay the foundation for further studies (e.g., primary school and academic preparation programs in upper secondary school), the use of AI should be determined by the tool's educational function. It is the individual's learning that is reflected in the guidelines for examination. On preparation day exam, all aids are allowed, but during the exam itself, students are not permitted to communicate with others or use generative AI. This is justified by the goal of ensuring the fairest and most equal exam setting and is a central goal of many countries' school systems. Consideration of the individual student's competence is also central for the final grades that teachers determine. In academic preparation programs, it is the individual student's knowledge and skills that are central: what the student is capable of without AI is still important.

**Learning mechanisms and limitations: A cognitive theoretical perspective**

Learning involves a change in knowledge through experience and takes place in the individual's brain. The process can start with attention to external stimuli such as words, images or movements that are sensed via sight, hearing, smell, taste and touch (Willingham, 2023). When we actively process these stimuli through thinking, repetition or reformulation, something happens with our memory: we may succeed in remembering. Whether this happens depends, among other things, on repetitions and mental processes. Motivation is central as the driving force for these types of cognitive processing.

The theory of cognitive processing emphasizes that people can only process a limited amount of information at the same time (perhaps 3–7 information units), but they can switch between tasks, as when people read a text on their mobile and occasionally glance at a news broadcast (Anderson, 2020). In other words, people have difficulties doing two things at the same time. However, automation eases the pressure in mental processing; someone can knit while watching a film because knitting is a largely automatic skill that requires little mental capacity. On the other hand, it would be challenging for a beginner to learn to knit while watching a movie because both activities require attention.

A central conclusion is that the human cognitive system has limited capacity to process potential stimuli. For teachers, it is important to recognize that digital distractions can easily hinder students' concentration and thus learning opportunities (OECD, 2024). On the other hand, a new study shows that impulsive students with concentration problems can benefit from AI tools in schoolwork (Klarin et al., 2024). This is explained by the fact that AI tools appear to make it easier for them to carry out certain tasks. If this is to work, these researchers believe that schools must have clear rules for how AI tools can and cannot be used. However, one study is not enough to draw a firm conclusion about the role AI can have in concentration. We need more research to better understand the desirable functions that AI can have in the learning process.

Attention is fragile, and the temptations of entertainment can disturb focus and cognitive processing (Elstad, 2008). This has implications for teachers' learning management. If students become used to alternating between school tasks and breaks with entertainment, this can establish itself as a habit and a desire to continue clicking when boredom arises. Over time, this can become an obstacle to persistence. For example, reading short texts at the expense of long texts could lead to superficial learning outcomes.

The working memory is a central part of the mental system. For example, if one is asked the name of the world's highest mountain, 'Mount Everest' will quickly be pulled from the long-term memory. This is an example of retrieval (or recall) from long-term memory to working memory. Long-term memory contains the residues of thinking ahead of time and consists of both explicit and implicit knowledge, which together form prerequisites and background knowledge (Willingham, 2023). The difference between working memory and long-term memory is essential; working memory is the place where thinking happens in the brain. In practice, several places in the brain are active at any given time, while long-term memory stores knowledge.

Working memory holds information that we actively need or use but, as noted above, it has limitations in terms of processing stimuli. However, working memory does not have limitations when retrieving the prerequisites stored in long-term memory, which makes that background knowledge a significant strength in the process of learning new material (Grissmer et al., 2023; Simonsmeier et al., 2019). This implies that students with adequate prerequisites have a clear advantage in the learning process and can more easily acquire new subject matter than those with fewer prerequisites. What constitutes adequate background knowledge will vary by subject and topic complexity, but in general, prerequisites include what we call fundamental concepts within a given field. Retrieval strengthens the memory itself.

The new topic should logically build on previous subject matter that has been covered in instruction, and students should have the necessary knowledge and skills required to work with the new topic. By ensuring that students have adequate prerequisites, teachers can facilitate smooth transitions to new topics and help students build on what they already know. The cognitive processing of prerequisites (along with the subject matter to which students are exposed) leads to generative learning, which involves actively creating meaning out of information to be learned by mentally reorganizing and integrating it with one's prerequisites, thus making it possible for one to use what has been learned in new situations (Fiorella & Mayer, 2016).

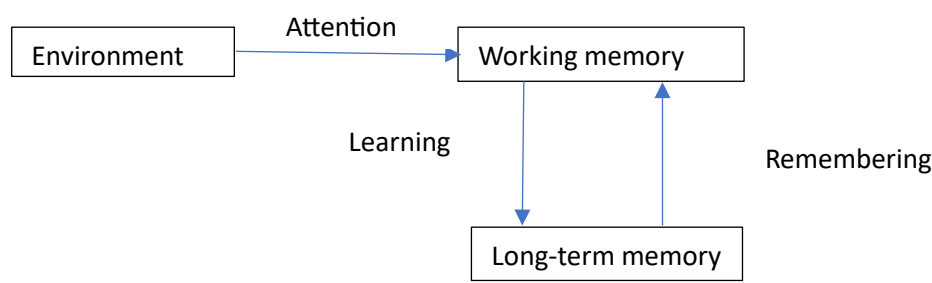

***Figure 1: A model of cognitive processing (Willingham, 2021).***

When stimuli processed in working memory are transferred to long-term memory, the new information is incorporated into existing schemata or forms new schemata. The knowledge is stored in a hierarchical manner. A schema is a set of thoughts about a topic that serves as a reference point for new information. Schemata make thinking faster, but when new information does not fit well with existing schemata, new ones must be formed. We must distinguish between long-term learning and the short-term learning of individual information elements. It is long-term learning that is central in grasping the material in school subjects, but new knowledge elements can be building blocks in a cumulative process. This means that new knowledge builds on previously learned knowledge and that the student can link knowledge elements not only within a single school subject but also themes and content that span multiple disciplines (Markauskaite et al., 2024). The concept of progression is central here; it refers to how instruction facilitates the development of students' knowledge and understanding over time – hour by hour, week by week and year by year. From such an outlook, learning becomes a cumulative and long-term pursuit (Willingham, 2023).

According to the cognitive theoretical perspective, memory consists of the residues of our previous thinking (Willingham, 2021). Thorough and deep thinking about school subject matter is more likely to leave a lasting impression in the memory than superficial processing of the same material. This assumption is in line with research in cognitive psychology, which shows that the more deeply and actively we engage with information, the more likely it is to be encoded into long-term memory (Anderson, 2020).

Above, we have noted that the student's cognitive capacity sets limits. However, the combination of visual and verbal information can improve learning effects without the combination itself contributing to exceeding processing capacity. Multimedia learning, which involves the simultaneous use of both text and visual materials, is effective because humans have separate information processing systems for visual and verbal information (Paivio, 2014). This allows larger amounts of subject matter to be processed at the same time and makes it easier to incorporate and understand information. Presenting subject matter through the two channels of visual and verbal can thus promote insight and deep learning, especially when that information is well integrated with students' prerequisites. Language models like ChatGPT present text in response to a prompt, but other forms of generative AI – such as like Midjourney, DALL-E, Canva and Beatoven – create images, videos, music and the like. It is possible that there will be tools for use in schools that combine verbal and visual messages. This could, for instance, occur within the framework of educational platforms like Curious, SchoolAI and Canvas. In that case, a new chapter in the development of tools will be written. The expected combination of words and pictures in a chatbot can make the subject matter easier to understand and strengthen learning.

**Desired difficulties in teaching to promote perseverance**

Does using AI in education make learners less intelligent? It depends. The question certainly requires a nuanced answer. The student's knowledge, which is stored in long-term memory, is of great importance for achieving reading comprehension with academic texts (Grissmer et al., 2023). On the other hand, we know that generative AI can act as a powerful tool that helps many who work with texts produce better work (Zhang & Noyes, 2023). Here, we must distinguish between machine performance and a student's learning outcome: AI tools can be a shortcut to producing text and images, music and the like but are not a shortcut for learning.

The goal of education is for students to remember what they have learned long after it has been taught. A related goal is to help students form a mental representation of a phenomenon that they can quickly and easily recall when necessary and apply in new situations. The latter process is called the transfer of knowledge. For students to be able to transfer knowledge, understanding the subject matter is essential. In the overarching part of the curriculum, emphasis is placed on deep learning, which highlights that students gradually deepen their insights into and understanding of concepts, procedures, and connections in school subjects (long-term learning) but also understand themes and questions that stretch across multiple disciplines. With deep learning, students engage in analytical thinking, problem-solving and reflection on their learning process to build a deep and lasting understanding. To truly master something and understand it in depth requires taking an active role in one's learning processes, using relevant learning strategies and the ability to evaluate one's own learning progress. This contributes to organized knowledge.

Variation theory (Marton, 2015) emphasizes students needing to see a pattern of variation: that is, different but specific ways to approach content. Learning is facilitated by looking at differences instead of similarities. This is considered an effective way to facilitate deep learning. In learning processes that are supposed to lead to understanding, a deep cognitive processing by solving tasks is therefore required. What role can AI play? It can supplement teachers by providing more examples and explanations than the teacher does in the lessons or explanations that the learning materials provide (Mollick & Mollick, 2023, 2024). This can be particularly useful outside the classroom, as with homework. The tasks that students are supposed to solve should be appropriately challenging; they should cause the desirable difficulties described above (Bjork & Bjork, 2020; Kirschner et al., 2022). In a didactic context, difficulties can be advantageous and therefore desired because they can stimulate coding and recall processes that enhance learning, understanding and memory. Therefore, teachers and textbook authors often create tasks that are intended to be appropriately demanding for the students, which means they contribute to progression. However, there are controversies regarding how teachers should best perform their role when students work on tasks that correspond with the idea of desirable difficulties: should the students explore and discover on their own (Hmelo et al., 2007), or should the teacher provide explicit explanations of the questions that the students have when they solve tasks, which is known as explicit teaching (Clarke et al., 2012; Kirschner et al., 2006)? Cognitive constructivism focuses on individuals actively constructing their knowledge. Piaget (1970) argued that premature teaching could hinder children from fully understanding through their own discovery. A theorist like Bruner (1961, 1966) recommended discovery learning, but several studies, such as those by Klahr and Nigam (2004) and Mayer (2004), found a weak learning effect of this exploratory method. Research indicates that we need to look at exploration as a teaching method in a nuanced way. The risk of overgeneralization is great. Yet, it must be assumed that a student's age and prerequisites, along with the nature of the school subject, matter for how favourable exploration is for learning progress. Older and more knowledgeable students can be expected to benefit more from exploration as a work process than younger students (Alfieri et al., 2011; Blanchard et al., 2010). It has been suggested that a combination of exploration with explicit teaching can be effective if the student has adequate prerequisites and sufficient motivation for schoolwork (Minner et al., 2010). Nevertheless, it is recommended that other teaching methods be used as well to leverage the advantages of various teaching methods (Sweller et al., 2023; Teig, 2023; Teig & Nilsen, 2022). The studies mentioned above do not include the use of AI. We are currently on uncertain ground regarding the interaction between students and AI tools. Until it is proven otherwise, I argue that a student's cognitive engagement, perseverance and propensity to delve more deeply are of great significance for learning progress when using AI tools in their learning work. Many teachers can relate to using dialogue as a tool to stimulate critical thinking and to achieve insight through questions and answers, the Socratic method; the teacher asks a series of well-considered fundamental questions to stimulate critical thinking, illuminate ideas or uncover assumptions that are not obvious. The purpose is to promote self-reflection, improve understanding and foster a more in-depth examination of ideas. This encourages participants to think critically and independently, which can lead to better insights. Through the process, students are stimulated to reflect on their viewpoints and assumptions, identify logical fallacies and develop the ability to argue logically.

A typical attribute of schooling is Socratic conversation: the teacher asks questions that might lead learners to a deeper level of ethical, philosophical, or moral thinking. This often involves the use of questions and hints to stimulate deep thinking and learning. It is possible to train some chatbots to conduct a conversation in the Socratic style. But teachers can also manually adjust the responses to include more inquisitive perspectives: they can ask the program to respond in a way that mimics Socrates's communication style in its dialogical and interrogative approach. More accurate instructions can improve the program's ability to give the desired responses. For example, 'Answer as Socrates by asking questions that encourage personal reflection and introspection'. Using the Socratic method in a conversation involves asking questions that lead to a deeper understanding. This can be achieved by posing open questions that promote reflection and analytical conversation.

Difficulties that make learning more challenging and extend the process of finding answers can result in deeper processing of information and better memory retention over time. These challenges are considered desirable difficulties because they trigger encoding and recall processes that lead to stronger learning outcomes and improve the long-term recall of the learning content (Bjork & Bjork, 2020). For learning to be effective, it must be an active process that should involve a change in the student's mental landscape. Good learning thus requires a cognitive change in the student, which can be achieved through the student's own actions– the activities in which they participate in. The teacher's role is relevant if it contributes to motivating students to participate in activities they would have otherwise avoided or simply ignored, but it is not limited to ensuring that students acquire information through reading texts created by AI, internet searches or even traditional paper books. If one accepts that overly easy access to an answer can deprive a student of a learning opportunity, part of the teacher's task is also to explain how students can use AI in ways that genuinely promote learning. For example, the unquestioned use of AI will mean that students deprive themselves of learning opportunities. Inappropriate use of AI can have both a gender-specific and social bias since students' abilities to persevere when working with educational material can manifest differently among boys and girls (Duckworth, 2017). It remains an open question whether there are gender differences in students' use of AI tools, but a study of higher education students indicates that there are (Carjaval et al., 2024). The extent to which parents support their children's schoolwork and express expectations also matter in terms of outcomes (Eriksson et al., 2021). For these reasons, it is important that schools help students avoid harming their own learning processes when using AI to take shortcuts. For example, teachers can emphasize how AI can be used as a tool for learning that supplements rather than replaces their own efforts.

The existential rationale for schooling is that students should develop their social intelligence. Learning can involve acquiring not only knowledge and skills but also attitudes and values that are relevant to a student's future. As Biesta (2015) notes, schools should ensure that students develop Bildung. It can involve the school's purpose concerning the development of the student's personality and character, intellectual capacity, moral views, and self-perception. The core of education is that the teaching process and learning resources contribute to forming responsible, autonomous individuals who can participate in a democratic society and acquire the necessary skills and knowledge for their future professional lives and personal development. What we do not know is how the AI phenomenon will impact students' formation and social intelligence.

Socratic questioning technique is built into certain conversational bots, such as Khanmigo (Khan, 2024a, 2024b), which covers subjects such as mathematics, humanities, coding and certain social sciences. Khanmigo does not just give answers but is designed to guide students to find the answer themselves. Khan (2024a) presents a vision of how Khanmigo can be used to improve teaching and assessments and complement what we have thus far known as classroom experiences, but we currently lack research on Khanmigo. We do not know what kind of evidence about the learning progress that students' interactions with Khanmigo lead to because the reports from these trials with Khanmigo and similar efforts attempts (Holt, 2024) are based on samples of fewer than 5% of students. This means that the evidence base is still too thin to draw firm conclusions. When we lack evidence, we must use our reason. An example of an evidence-based AI tool can be found in an application of GPT Tutor (Bastani et al., 2024). The student asks for the answer directly, but the chatbot will only provide the student with hints on how to proceed.

This is an example of how a chatbot can be preset to offer hints instead of giving the student the answer itself, thus ensuring a desirable difficulty in the learning process. In mathematics, this type of personalized guidance tool can be beneficial for learning because it gives the student something to ponder instead of simply providing the answer (Bastani et al., 2024). Whether this conclusion is valid across other contexts is as yet unknown.

**Motivation**

Motivation is the force that drives a person to act, and motivation encompasses both thoughts (such as a goal expressed as a desire) and feelings. Thoughts and feelings are often mixed. Intrinsic motivation stems from the interest in the activity itself and provides satisfaction when performing it, while extrinsic motivation is driven by external rewards or goals that a student sets. Often, intrinsic and extrinsic motivation are mixed.

A plausible motivation mechanism is motivation -> effort -> learning. But this may be a false overgeneralization. Although motivating students is important, it is not necessarily intrinsic motivation that leads to learning. Succeeding at a task provides a positive experience, which releases chemicals that transmit signals between nerve cells in the nervous system; at least one of these – dopamine – is linked to reward, motivation and pleasure. Effort and perseverance are central to success in school: being motivated is not the same as simply feeling a desire for something. It is a betrayal to suggest to young people that learning can always be fun and enjoyable, and it gives a misleading impression that learning can be acquired without effort (Haake & Gulz, 2024. p. 111). Students' motivation, self-discipline and the framework around the teaching situations (what is called learning leadership above) are important for learning. Most often, these internal and external factors are mixed in school contexts, and effort is typically accompanied by negative feelings (David et al., 2024). Therefore, maintaining motivation can be challenging for students struggling with endurance and exercising self-control. The concept of self-discipline represents the will to persist when faced with the desire to learn. Perseverance is not innate; students must train themselves to regulate their own attention ability, and school is a suitable arena for this important effort.

Research shows that self-discipline affects learning outcomes more than intelligence (Duckworth & Seligman, 2005), and grit (that is, perseverance and commitment over time) is important for schoolwork (Duckworth, 2016). This resilience can be developed through individual choices and environmental influences and is associated with a growth mindset, where the student believes learning ability can be improved with effort (Dweck, 2015).

What role does AI play in and out of the classroom in this regard? Research so far does not provide a clear picture. When chatbots give a direct answer to tasks, there arises a temptation for the student to copy the answer without engaging in the task. This weakens the learning process. This point can be illustrated by an experiment involving university students. A researcher divided the class into three groups and assigned them to solve tasks in two phases (Shein, 2024). In the first phase, the researcher divided the students into three groups. One group had access to ChatGPT, another had access to Llama (a slightly simpler language model than ChatGPT), and the third group could only use Google's search engine without access to AI. The results showed that the group using ChatGPT solved the task fastest. The group with Llama took a bit more time, and the group that only used the search engine took the most time. The reason was that the third group broke down the task into smaller components to solve the task step by step, which is a time-consuming process. The analysis of the process showed that Llama required more from the students than ChatGPT. In the second phase, the students had to solve the same task without aids of any kind. In other words, the second phase was a kind of exam situation. Here, the results were completely reversed. Those who had used ChatGPT remembered very little, and all students who participated in this ChatGPT group in the first phase failed the test. Half of those who used Llama passed the test, while all who had used only the search engine passed. This conclusion is consistent with the assumption of emphasizing desirable difficulties in the learning process: it is not favourable for the learning process if learners come too easily to the answers. Good learning requires effort, persistence and will. Another study examined the use of ChatGPT among university students (Abbas et al., 2024). The results from that study suggest that students who experience high workloads and time pressures tend to take shortcuts by using ChatGPT in ways that are not conducive to learning. It also showed that increased use of ChatGPT could lead to self-discipline problems in some students and negatively affect academic performance.

The conclusion we must draw is that research does not give us a clear and unequivocal answer to the question of which groups of learners can benefit from AI and which groups do not benefit from it. Perhaps we need to nuance the conclusions we draw that this may present itself differently in different subjects and at different age levels? This showed that frequent use of ChatGPT correlated with a tendency to procrastinate on assigned learning activities; the first led to forgetfulness and poorer academic performance. The authors believe that students should be encouraged to use ChatGPT as a supplementary learning tool rather than a shortcut to academic work and find a balance between support and personal effort. Therefore, teachers' learning leadership could be of great importance.

It is advantageous for the learning process that students engage in independent thinking and check their own knowledge before seeking answers from external sources such as AI tools or the internet (Giebl et al., 2022). Finding answers through AI tools or by through Google can indeed undermine the learning process. Therefore, teachers can motivate students to reflect on the material on their own before seeking information elsewhere. Independent thinking relies on mechanisms that activate existing prerequisites, awaken curiosity and metacognitive self-reflection and create deeper processing of the instructional material.

**Generative AI as an information retrieval tool**

AI based on LLMs is not an information retrieval tool per se. Nevertheless, AI can be used as a tool for students in their quest for information. If they encounter a difficult word (such as 'alienation'), AI can be used just as well as traditional search engines. ChatGPT, Gemini and Bing thus appear as alternatives to conventional search engines. Although traditional search engines often lead to useful websites, AI tools like ChatGPT can answer more complex questions that are not necessarily adequately covered in online encyclopaedias. For example, ChatGPT can provide concise answers on the similarities between classical Greek tragedies and Ibsen's play *Ghosts*. Even if the answer may be superficial, it gives the student a starting point for further investigation. Chatbots can offer ideas that the student can explore further in other literature, but this appears somewhat unfinished with respect to some cultural areas. Therefore, teachers and librarians should still play a role in advising students on where to find good sources that provide in-depth explanations.

ChatGPT can also adapt explanations of complex ideas to a more comprehensible level than online encyclopaedias, as when the conversation robot explains lightning to an eight-year-old. This kind of tailored explanation can be particularly useful for students with special needs or who have another mother tongue, as AI can provide explanations in languages in which students are proficient. This can make AI tools particularly useful for learning concepts. Although chatbots generally provide basic explanations, those are at least a foundational starting point for further learning and exploration. If one asks for more detailed answers, one gets more text.

AI tools do not think, but they do create human-like responses based on probabilities. The type of AI that exists today does not have the ability to truly understand the meaning behind the words that AI creates. This means that all answers produced by chatbots lack depth and insight to some extent. However, the learning individual must know what to look for and be able to assess the information found based on existing knowledge (Neelen & Kirschner, 2020, p. 130). Critical proficiency is more important than ever, even with access to AI and the internet.

**Generative AI in the classroom ecosystem**

Walther Doyle (2013) considers the classroom an ecosystem in the sense that there is a complex interplay between teacher, students, and digital devices. What typically happens in a classroom are variations between structure and flexibility, discipline and independence and individual and collective learning. This ecological system is complex and dynamic, and its properties are important for teachers to be able to deliver good instruction that can meet their students' needs. For this to be possible, teachers must think about how tasks are structured, how access to technology and the social dimensions in the classroom support or hinder learning and how the potential in the interactions is used. All these components must work together to create a suitably challenging learning environment. The teacher can play the roles of both the classroom's clear leader and of the mentor who assists the student(s) when help is needed. Activities will proceed in different phases in which the roles of teacher and students can change. From an ecological perspective, introducing AI into the classroom represents an additional component to the school ecosystem that can change the existing balance and interactions in the classroom.

The theory of distributed cognition looks at the interplay between learning and cognitive processes in learning individuals and technologies like AI. How can AI affect the interaction between teachers, students and learning resources in the classroom ecosystem? So far, we lack research to provide a thorough answer. However, time for individual follow-up of each student is a scarce resource in any classroom; instead of waiting for a teacher to have time to answer, students can ask questions to a chatbot. That tool may not always provide the correct answer, but it can in many cases provide usable information about students' questions. When technology takes over some of the tasks that were previously the responsibility of the teacher, the balance in the classroom's ecosystem may change. Some might argue that access to an AI resource can help redefine the teacher's role (Koh et al., 2023) from being the sage on the stage to the guide at the side. However, it is difficult to say that there is solid research-based evidence that this is beneficial for learning progress (Sweller et al., 2023). So far, we must say that the impact of AI tools on how teachers can and should carry out their role is unresolved.

AI tools are likely to only get better and even more widespread. Doyle's ecological theory, in combination with viewing students as intentional actors (Kirschner, 2022), can help teachers in designing teaching and assessment situations that are favourable for learning. An important premise is that students have incentives to use learning resources such as generative AI in ways that promote learning. Even with the incentives of which students are aware, some students may avoid using learning resources in ways that promote learning (Elstad, 2016). In such cases, a teacher who sets boundaries for activities by exercising the authority necessary is needed. AI challenges the teacher's **role**, but it is too early to say whether the teacher's role will change.

The teacher's authority defines the space of possibility for leadership in the classroom and can be divided into several categories. First, the teacher has professional authority based on the necessary formal competence attained through relevant education. Secondly, the teacher gains an institutional authority in which this professional competence is crucial for exercising the teaching role. The entire functioning of the school can contribute to the teacher's authority. Additionally, the individual teacher's personal authority plays a significant role. A teacher can build up a position of authority, for example, by adopting a proactive approach to teaching (Kounin, 1977), maintaining a steady flow of activities in teaching and protecting their efforts against interruptions and disturbances. However, a teacher can also lose personal authority.

**Learning with and without AI – Some preliminary conclusions**

When considering the potential benefits and challenges of integrating AI into the learning process, several questions require careful consideration. Learning with and without AI both have strengths and weaknesses, a deeper understanding of which can help teachers make more informed decisions about how best to use technology in education. AI does not solve the challenge that the content in school subjects is hierarchically organized while the student will read the chatbot's messages linearly. If the student simply copies answers from the chatbot, no learning occurs. The critical question is how thoroughly the student works to rebuild the hierarchical organization on which the message is based. Learning with understanding requires the student's full attention. The student needs material to think with, and a chatbot can provide such material. At the same time, the more the student knows, the deeper they can think. Therefore, the answer to how beneficial AI is for the learning process depends on how the student interacts with the chatbot's messages.

Over time, research will likely clarify when and under what conditions the use of AI tools is advantageous for the learning process. Individual teachers must weigh the pros and cons against one another and especially the students they have, the subject matter being discussed, the goals teaching is intended to achieve and so on.

Hodges and Kirschner (2024) suggest these pieces of advice to teachers who use AI. First, students' focus should be shifted from grade awareness to the learning process itself. When teachers can assess the actual work process of their students, they can better understand their learning progress. There are educational portals that make this possible. But such monitoring, combined with giving feedback, is obviously intensive in terms of time and labour. Second, technological development over time can alleviate the teacher's workload, so it may be wise to place greater emphasis on oral presentations of written work processes. This provides an incentive for students to remember and understand the subject matter with which they are working. Third, teachers can use AI tools to tailor assignments to be more specific, personal or context-dependent, such as tasks that are directly related to recent class discussions, current events or unique scenarios for which AI is less likely to succeed in generating answers. When teachers give assignments that ask students to reflect on their personal experiences or opinions, there is reason to assume that students will be less inclined to take shortcuts with AI-generated text. Fourth, the teacher should raise topics such as the ethical use of AI, including discussions about academic integrity, the limitations of AI, and how important it is for students to submit their own work. By using a combination of these strategies, teachers can try to mitigate the challenges of students taking tasks too lightly when they have access to AI tools.